\documentclass[pss]{wiley2sp} % provides new 2008 pss two-column style (no alternative manuscript style output available at present)
\usepackage{amsmath}
\usepackage{bm}              % uncomment these two packages if you
\usepackage{w-greek}         % need extended greek-letter functionality in math mode

\begin{document}

% Title of the article
\title{Quantum criticality of Ce$_{1-x}$La$_{x}$Ru$_{2}$Si$_{2}$ : the magnetically ordered phase}

% Abbreviated title for the page headers
\titlerunning{Quantum criticality of Ce$_{1-x}$La$_{x}$Ru$_{2}$Si$_{2}$}

% Authors
\author{%
  St\'ephane Raymond\textsuperscript{\Ast,\textsf{\bfseries 1}},
  William Knafo\textsuperscript{\textsf{\bfseries 2}},
  Jacques Flouquet\textsuperscript{\textsf{\bfseries 1}},
  Fr\'ed\'eric Bourdarot\textsuperscript{\textsf{\bfseries 1}},
  Pascal Lejay\textsuperscript{\textsf{\bfseries 3}}}

% Abbreviated list of authors for the page headers
\authorrunning{S. Raymond et al.}

%E-mail-address of corresponding author
\mail{e-mail
  \textsf{raymond@ill.fr}, Phone:
  +33 4 38 78 37 38, Fax: +33 4 38 78 50 98}

% author's affiliations/addresses
\institute{%
  \textsuperscript{1}\,CEA-Grenoble, INAC / SPSMS, 38054 Grenoble Cedex, France\\
  \textsuperscript{2}\, 
LNCMI, 31400 Toulouse, Cedex, France\\
 \textsuperscript{3}\,Institut N\'eel, CNRS, 38042 Grenoble Cedex, France}

\received{XXXX, revised XXXX, accepted XXXX} % do not change, will be filled in by the publisher
\published{XXXX} % do not change, will be filled in by the publisher

%Please select four to six PACS-codes from the enclosed list (PACS.txt) or from www.aip.org/pacs)
\pacs{ } % For example: 71.20.Ps

\abstract{%
% This is a macro for the typesetting of two-column text in an
% abstract. It will typeset the two arguments in \abstcol{}{} as the
% left and right column inside the abstract box. At the
% columnbreak there will be always a columnbreak (\par), so both
% columns start with a new paragraph. No automatic column height
% balancing is done.
%
% If used with a \titlefigure it will silently output both
% parameters as consecutive paragraphs.
%
% The macro is defined exclusively inside the argument of \abstract{};
% if used outside it will raise an error.
%
% Usage: \abstcol{<left column>}{<right column>}
We report specific heat and neutron scattering experiments performed on the system Ce$_{1-x}$La$_{x}$Ru$_{2}$Si$_{2}$ on the magnetic side of its quantum critical phase diagram. The Kondo temperature does not vanish at the quantum phase transition and elastic scattering indicates a gradual localisation of the magnetism when $x$ increases in the ordered phase.}

% The class file requires the standard graphicx Latex package. See the 'LaTeX
% standard graphics and color packages documentation' for more information at
% <http://tug.ctan.org/tex-archive/macros/latex/required/graphics/grfguide.pdf>.
%
% Accepted figure file formats depend on which LaTeX flavour is used.
% Classic LaTeX is always able to use Encapsulted Postscript (EPS);
% PDFLaTeX can't use this but accepts PDF, JPG, PNG, and GIF formats.
%
% See examples for implementing graphics in floating figure environments later in this file.
% If \titlefigure is given, it takes as its mandatory parameter the
% name (without extension) of some figure file.
%\titlefigure[height=3.1cm]{empty2w}
%\titlefigurecaption{%
  %This is the caption of the \emph{optional} abstract figure. If
  %there is no abstract figure here, the abstract text should be divided into both columns.}

\maketitle   % please do not remove

One important issue in the study of quantum criticality in itinerant electron systems is the determination of the dynamical spin susceptibility and of its relation to the anomalous bulk properties observed near the quantum critical point (QCP). To this aim, the description of 3d itinerant electron systems was achieved by the spin fluctuation theory of Moriya. In 4f electron systems,  the occurence of  the Kondo effect with a related energy scale of order 10 K leads to  a more complex problem. Basically, two scenarii emerge for these systems: the conventional spin fluctuation scenario in the line of Moriya's theory \cite{Moriya,Hertz,Millis} and the local scenario \cite{Coleman}. Their respective description of the intrinsic nature of quantum criticality deeply differ.  In the spin-density wave scenario, the fluctuations of the order parameter, i.e. at $\textit{the wave-vector of ordering}$, are leading to the transition as in a classical phase transition. A difference with a finite temperature phase transition is that the effective dimension of the system is increased due to quantum fluctuations \cite{Hertz,Millis}. In this theory, the Kondo effect evolves smoothly accross the QCP. On the contrary, in the so-called local scenario, the low frequency spin dynamics is critical $\it{everywhere}$ in the Brillouin zone. In this model, this is associated with a destruction of the Kondo effect  at the QCP. 

Over several decades, numerous neutron scattering experiments were performed on the archetypal Pauli paramagnet heavy fermion compound CeRu$_{2}$Si$_{2}$ for various dopings leading to long range magnetic ordering \cite{Kambe,Knafo2,Kadowaki,Raymond}. For Ce$_{1-x}$La$_{x}$Ru$_{2}$Si$_{2}$, sine-wave modulated magnetic ordering occurs for $x$ $\ge$ $x_c$=0.075 at the incommensurate wavevector $\bf{k_1}$=(0.31, 0, 0), the magnetic moments being aligned along the c-axis of the tetragonal structure, due to the strong Ising anisotropy originating from the crystal field. In the present paper, we report recent data taken in the magnetically ordered phase for $x$=0.13 ($T_N$=4.4 K) and $x$=0.2 ($T_N$=5.8 K) that extends the works performed in the past for $x$=0.13 \cite{toto}, $x$=0.2 \cite{Regnault,Jacoud} and $x$=0.3 \cite{Mignot}.

Figure 1 shows the magnetic specific heat divided by the temperature for several concentrations accross the critical point (data for $x$=0 and $x$=0.075 are taken from Ref.\cite{Kambe}).
While focus is usually made on the critical concentration, the most striking features occur in the ordered phase.
For x=0.13, $C_m/T$ is almost constant in the ordered phase. For $x$=0.2, the slow decrease of $C_m/T$ below $T_{N}$ is amplified below a second transition temperature $T_{L}$ = 1.8 K. For $x$=0.13, the low transition occurs at $T_L$ $\approx$ 600 mK. We note that the decrease of $C_m/T$  below $T_L$ for $x$=0.13 is far smaller than the one reported earlier \cite{Fisher}. The huge effective mass related to the high value of $C_{m}/T$ for T $\rightarrow$ 0 at $x$=0.13 already suggests that itinerant magnetism is characteristic of the ordered phase. The anomalous shape of the $x$=0.13 specific heat curve can be qualitatively understood as a the sum of a large spin fluctuation contribution and of a specific heat jump related to the magnetic phase transition. To better understand the magnetic state of Ce$_{1-x}$La$_{x}$Ru$_{2}$Si$_{2}$, neutron scattering measurements were undertaken for $x$=0.13 and $x$=0.2 on the cold three axis spectrometer IN12 at the Institut Laue Langevin, Grenoble. Details of the inelastic measurements are given elsewhere \cite{Knafo1}.

%Please use for changes during revision the following colour change option:
%\begin{changed}
  %This is a text snippet marked as \emph{changed}.
  %This is done by enclosing it in an environment called \verb+changed+. Please note
  %that in certain circumstances there might be small side effects such
  %as make up deviations or additional blanks.
%\end{changed}

\begin{figure}[t]%
%\vspace{-1cm}
%\hspace{1cm}
\includegraphics*[width=7cm,height=7cm]{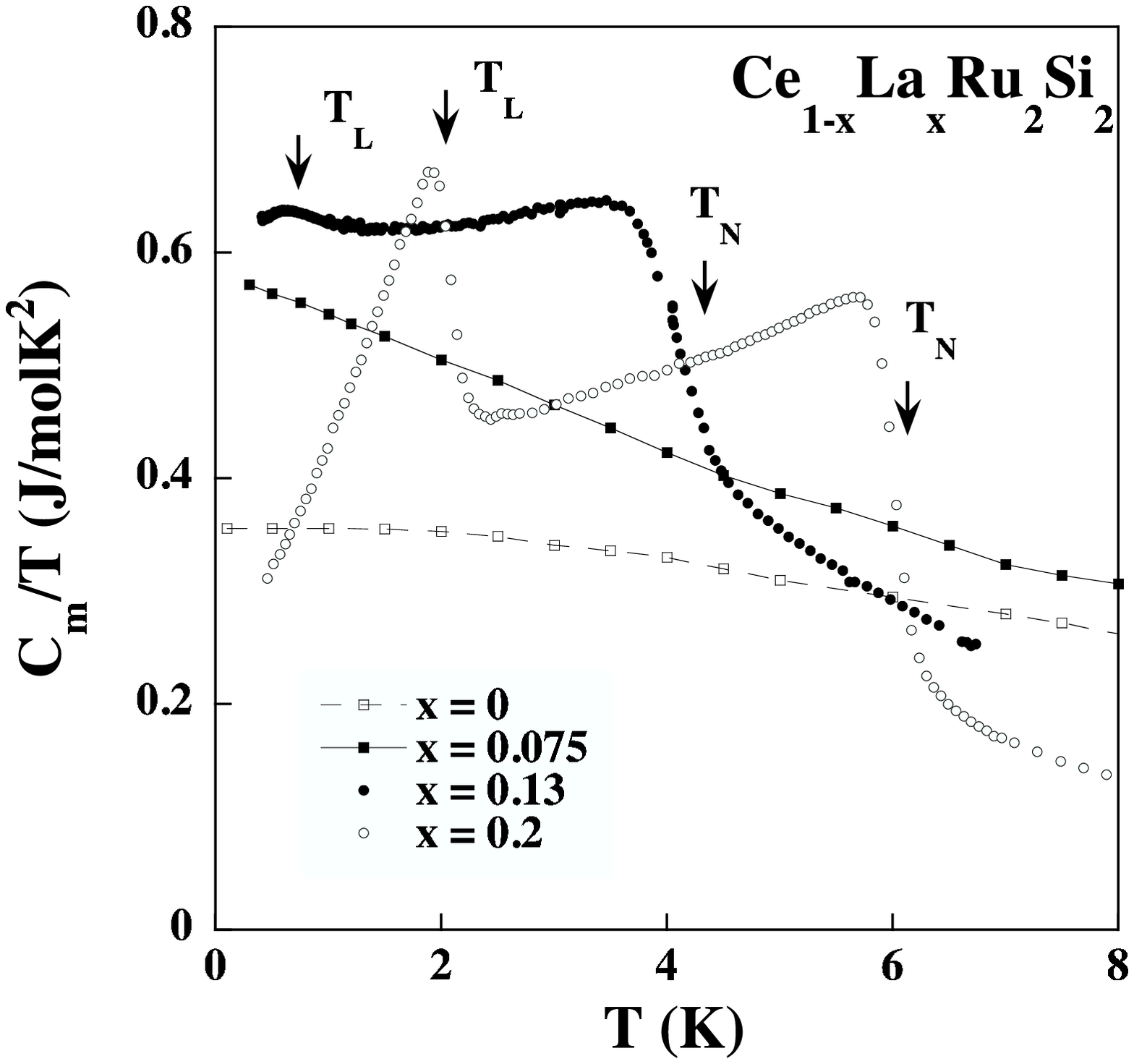}
\caption{Magnetic specific heat of Ce$_{1-x}$La$_{x}$Ru$_{2}$Si$_{2}$ divided by temperature for $x$=0, 0.075, 0.13 and 0.2.}
%\vspace{-0.5cm}
\label{onecolumnfigure}
\end{figure}

\begin{figure}[t]%
\includegraphics*[width=8cm,height=8cm]{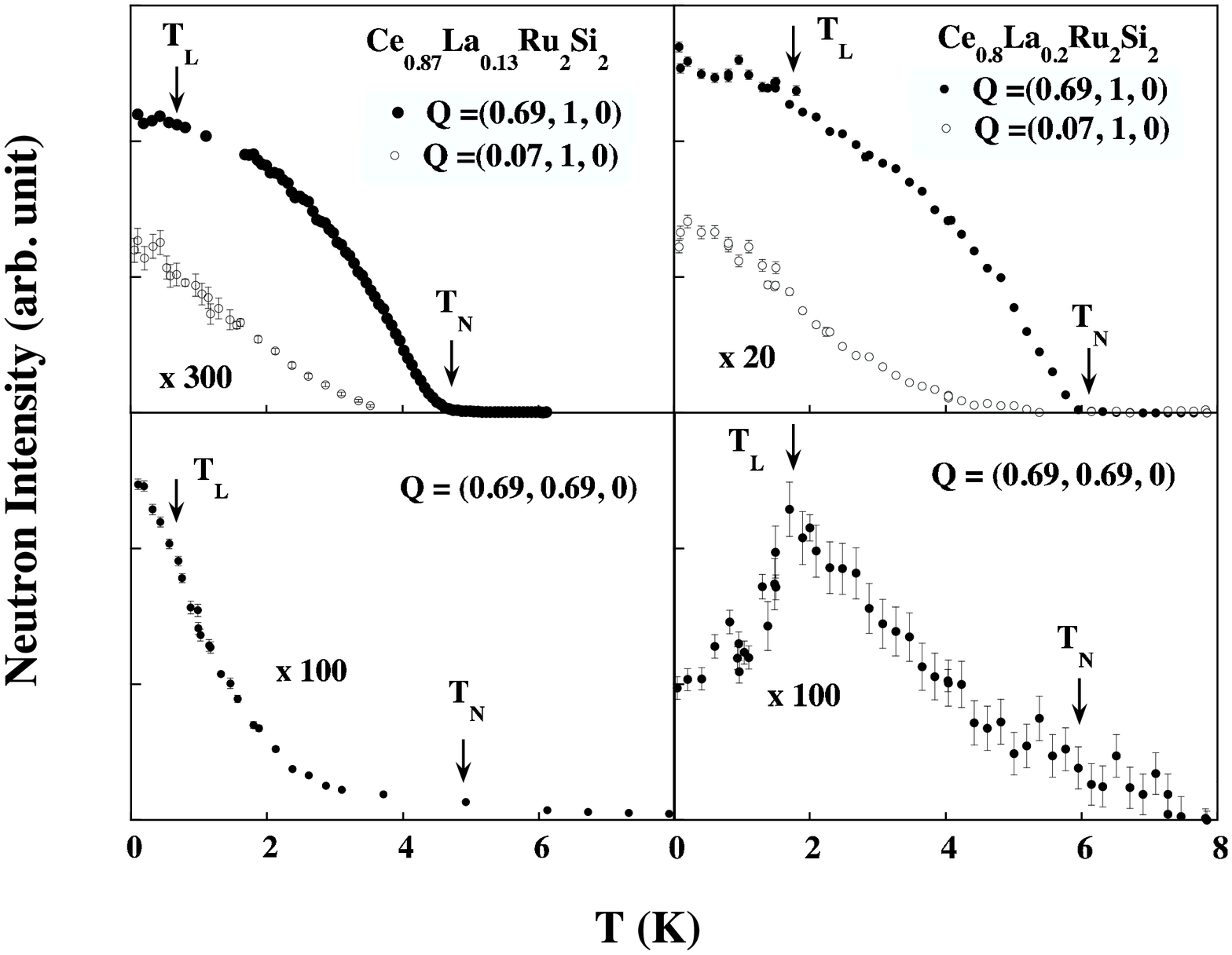}
\caption{%
 Temperature dependence of the elastic scattering measured at $\bf{Q}$=(0.69, 1, 0), $\bf{Q}$=(0.07, 1, 0) and $\bf{Q}$=(0.69, 0.69, 0) for x=0.13 (left) and x=0.2 (right).}
%\vspace{-0.7cm}
 \label{onecolumnfigure}
\end{figure}

\begin{figure}[t]%
\includegraphics*[width=8cm,height=4.7cm]{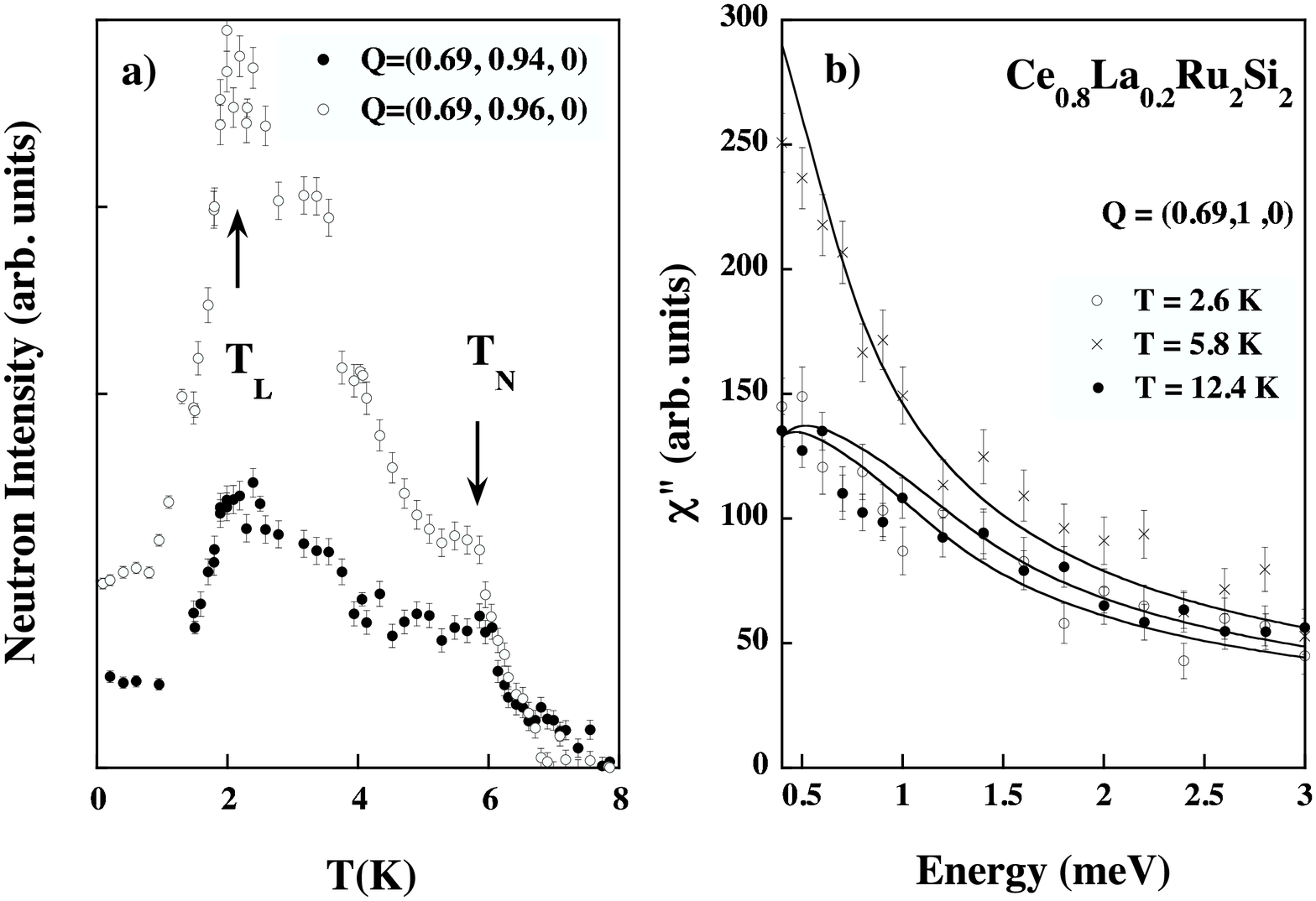}
\caption{a) Temperature dependence of the diffuse scattering measured at $\bf{Q}$=(0.69, 0.96, 0) and $\bf{Q}$=(0.69, 0.94, 0) for $x$=0.2. b) Energy spectra measured for $\bf{Q}$=(0.69, 1, 0) at 2.6, 5.8 and 12.4 K.}
%\vspace{-0.5cm}
\label{onecolumnfigure}
\end{figure}
The temperature dependence of the elastic intensity is shown in figure 2 for the wave-vector of ordering $\bf{k_1}$=(0.31, 0, 0) (measured at the momentum transfer $\bf{Q_1}$=(0.69, 1, 0)=(1, 1, 0)-$\bf{k_1}$), the third harmonic of $\bf{k_1}$ (measured at the momentum transfer $\bf{Q}$=(0.07, 1, 0)=(1, 1, 0)-3$\bf{k_1}$), and for the wavevector $\bf{k_2}$=(0.31, 0.31, 0) at which short range magnetic fluctuations are observed in pure CeRu$_{2}$Si$_{2}$ and here conveniently measured at $\bf{Q}$=(0.69, 0.69, 0)=(1, 1, 0)-$\bf{k_2}$. For $x$=0.13, the third harmonics is barelly developped with $I_{\bf{k_1}}$/$I_{\bf{3k_1}}$ $\approx$ 500, while fluctuations at $\bf{k_2}$ are substantial with $I_{\bf{k_1}}$/$I_{\bf{k_2}}$ $\approx$ 100 ($I_{\bf{k}}$ is the elastic intensity at the wavevector $\bf{k}$). For $x$=0.2, the elastic scattering at $\bf{k_2}$ is maximum at $T_L$ and  $I_{\bf{k_1}}$/$I_{\bf{k_2}}$ $\approx$ 300 at 50 mK. Similar maxima of the elastic scattering at $T_{L}$ are observed at wave-vectors in the vicinity of $\bf{k_1}$. This is shown in figure 3a for $\bf{Q}$=(0.69, 0.96, 0) and $\bf{Q}$=(0.69, 0.94, 0). 
For these wave-vectors "in the foot" of the $\bf{k_{1}}$ peak (the corresponding full width at half maximum of the magnetic Bragg peak being 0.04 r.l.u.), a maximum of the diffuse scattering is expected at $T_{N}$ for a classical finite temperature phase transition.
Such diffuse scattering in the vicinity of $\bf{k_1}$ is not observed for $x$=0.13. Concerning the third harmonics, we obtained $I_{\bf{k_1}}$/$I_{\bf{3k_1}}$ $\approx$ 40 for $x$=0.2. This increase of the third harmonics from $x$=0.13 to $x$=0.2 points out to an increase of the local nature of the magnetism. Indeed it corresponds to a stronger squaring of the modulated structure and consequently to a decrease of the longitudinal fluctuations ($I_{\bf{k_1}}$/$I_{\bf{3k_1}}$=9 for a squared modulation). We stress on the fact that the nature of the magnetic ordering below $T_{L}$ is still unknown. Current hypotheses correspond to a lock-in of the phase of the modulation \cite{Regnault} or to a transition from a single-$\bf{k}$ structure to a double-$\bf{k}$ structure \cite{Jacoud2}.
\begin{figure}[t]%
%\hspace{1cm}
\includegraphics*[width=7cm,height=7cm]{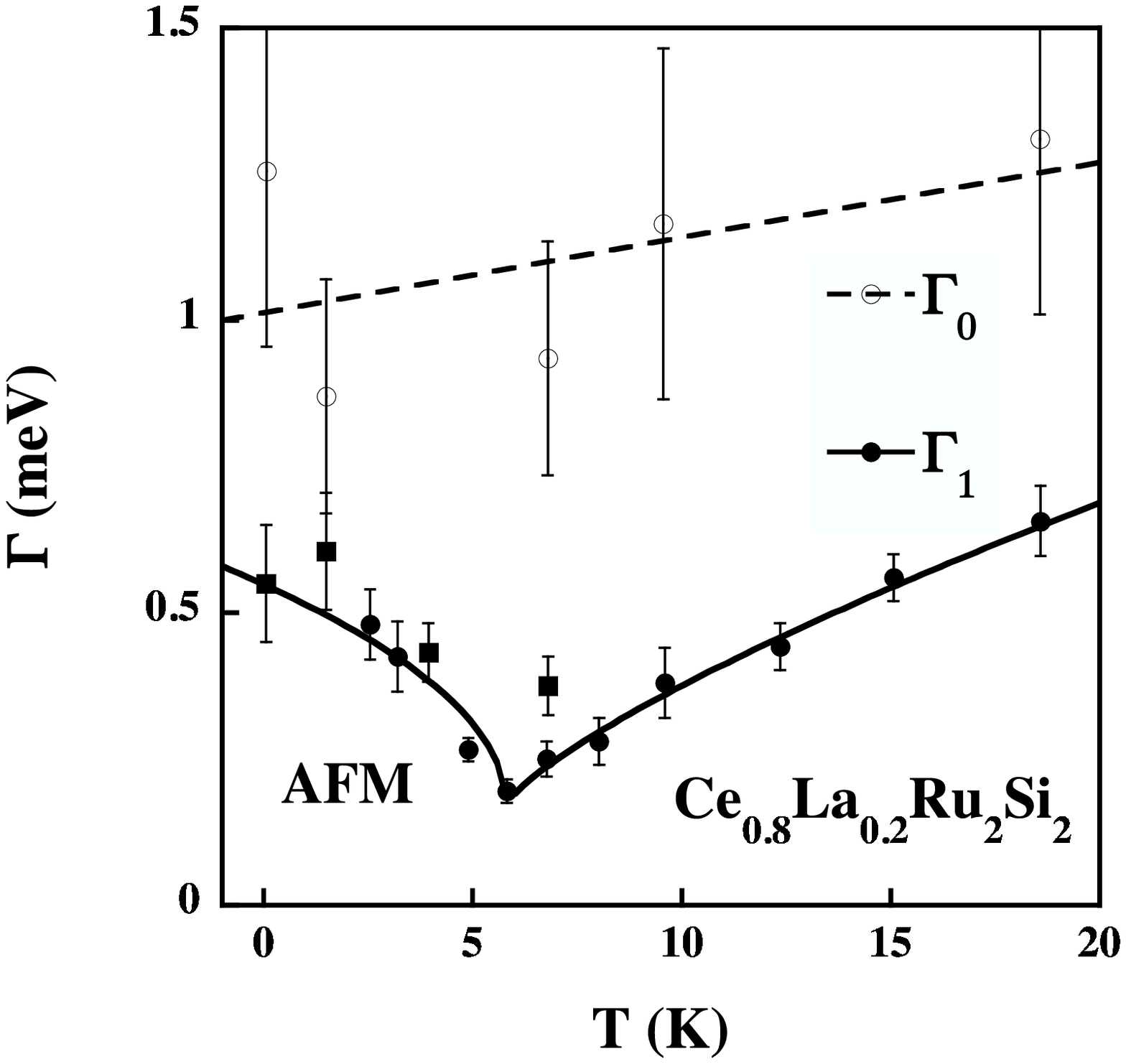}
\caption{Local (open symbols) and antiferromagnetic (full symbols) relaxation rate of Ce$_{0.8}$La$_{0.2}$Ru$_{2}$Si$_{2}$ as a function of temperature. Full circles correspond to measurements performed at $\textbf{Q}$=(0.69, 1, 0) while full squares correspond to measurements performed for $\textbf{Q}$=(0.72, 1, 0) i.e. slightly off the $\bf{k_1}$ peak position. The local relaxation was measured at $\bf{Q}$=(0.44, 1, 0).}
\label{onecolumnfigure}
\end{figure}
Figure 3b shows typical energy spectra obtained for $x$=0.2 at the constant momentum transfer $\bf{Q}$=(0.69, 1, 0) corresponding to the wave-vector $\bf{k_1}$ for $T$=2.6, 5.8 and 12.4 K (See Ref.\cite{Knafo1} for more details). Such spectra were analyzed using a Lorentzian lineshape characterized by a single linewidth corresponding to a wave-vector dependent relaxation rate. Figure 4 shows the temperature dependence of the magnetic relaxation rates measured by inelastic neutron scattering for $x$=0.2 at two wave-vectors \cite{Knafo1}. At the wave-vector $\bf{k_1}$ of magnetic ordering, the relaxation rate $\Gamma_{1}$ is minimum at $T_{N}$. However its value is finite as for the value of the relaxation rate at the critical concentration $x_c$ for T $\rightarrow$ 0 \cite{Knafo2} (See below). The relaxation rate $\Gamma_0$ corresponding to fluctuations measured far from the ordering vector, here at the momentum transfer $\bf{Q}$=(0.44, 1, 0), does not show any anomaly at $T_{N}$. The persistence of local relaxation inside the magnetically ordered phase is interpreted in terms of a finite Kondo temperature at $T$ $\rightarrow$ 0 for $x$=0.2. Very similar data were also obtained for $x$=0.13 \cite{Knafo1}. 

\begin{figure}[t]%
%\hspace{1cm}
\includegraphics*[width=7cm,height=7cm]{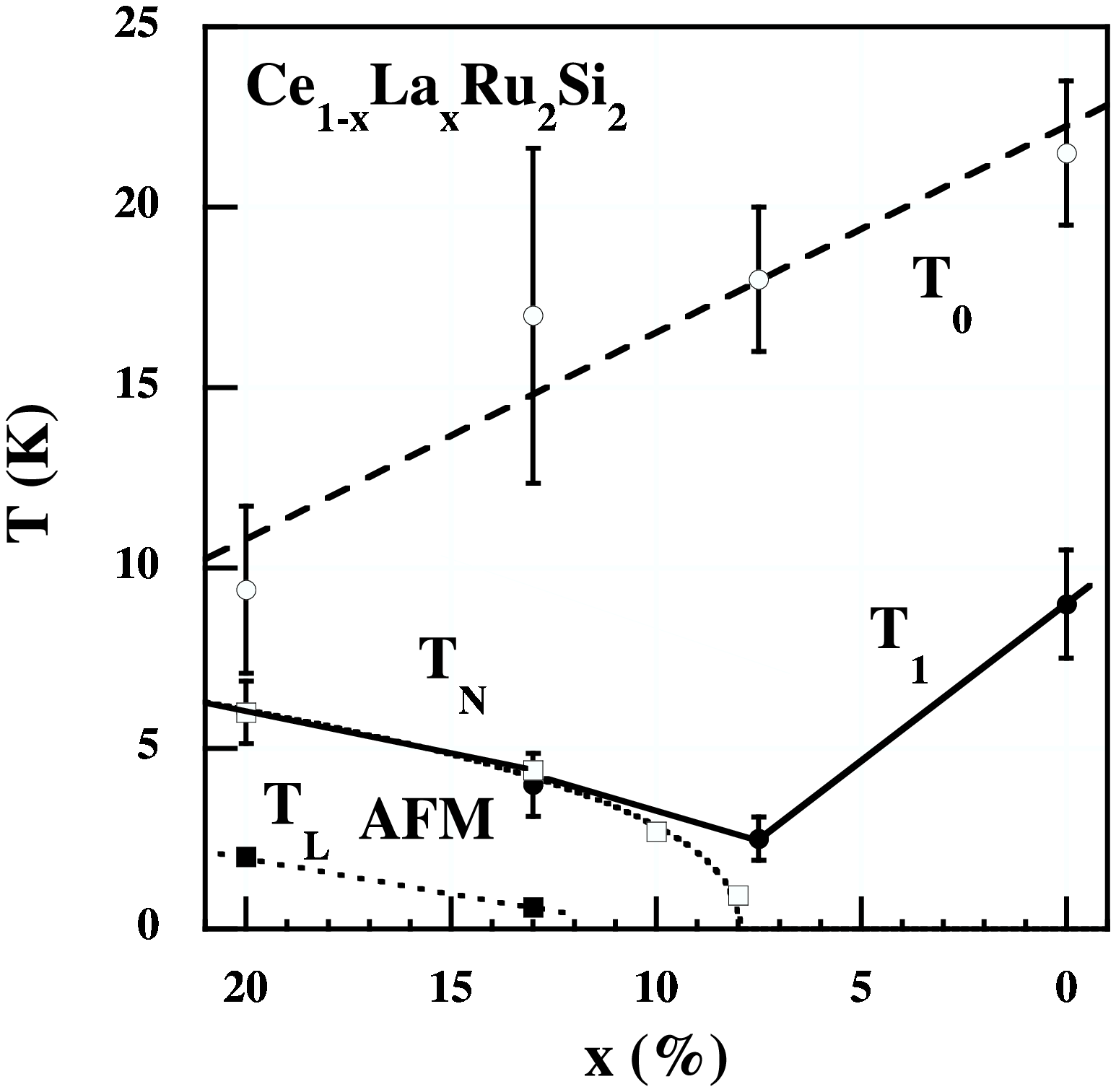}
\caption{Magnetic phase diagram of Ce$_{1-x}$La$_{x}$Ru$_{2}$Si$_{2}$ obtained by elastic and inelastic neutron scattering \cite{Knafo1}.}
\label{onecolumnfigure}
\end{figure}

Figure 5 summarizes the quantum critical phase diagram obtained by neutron scattering measurements \cite{Knafo1}. $T_{1}$ is the extrapolation of the relaxation rate $\Gamma_1$ for $T$ $\rightarrow$ 0 ($k_{B}T_{1}=\Gamma_{1}(T)_{T \rightarrow 0}$) and $T_{0}$ is the extrapolation of $\Gamma _0$  for $T$ $\rightarrow$ 0 ($k_{B}T_{0}=\Gamma_{0}(T)_{T \rightarrow 0}$). $T_{1}$ shows a minimum at $x_c$ and is finite \cite{Knafo2}. $T_{0}$ does not show any anomaly at $x_c$, which evidences again that the Kondo fluctuations are finite in the magnetically ordered phase. It is worthwhile to note that $T_1$=$T_N$ in the ordered phase. Figures 4 and 5 show that the behavior at $T$ $\rightarrow$ 0  as a function of $x$ is similar that the one for $x$ =0.2 as a function of $T$. The minimum of $T_{1}$ at $x_{c}$ (or of $\Gamma_{1}$ at $T_{N}$) and the absence of anomaly of $T_{0}$ at $x_c$ (or of $\Gamma_0$ at $T_{N}$) suggest that the fluctuations at the wave-vector $\bf{k_1}$ are responsible for both the classical and quantum phase transitions. The finite value of the characteristic energy at $T_{N}$ or $x_{c}$ suggests a first order character of the phase transition.

To conclude, our data confirm that the quantum critical point of Ce$_{1-x}$La$_{x}$Ru$_{2}$Si$_{2}$ is of itinerant nature.
This is evidenced by the finite value of the Kondo temperature in the ordered phase and by the associated absence of anomaly in the local relaxation rate at $x_{c}$ or $T_{N}$.
In the magnetically ordered phase, the magnetism gradually evolves from itinerant to localized, as evidence by the increase of the squaring of the modulation and by the maximum of diffuse scattering occurring at the  finite temperature $T_{L}$ for $x$=0.2 but absent for $x$=0.13. The fact that it occurs at $T_{L}$ rather than at $T_{N}$ as expected is unclear and would merit further investigation, in relation with the question of the nature of the second transition at $T_{L}$.

%\begin{acknowledgement}
%An acknowledgement may be placed at the end of the~article.
%\end{acknowledgement}

% Use the following code if you wish to generate your bibliography with BibTeX;
% replace the string "pss-demo" below with the name(s) of
% the BibTeX data base(s) you want to use.
% The resulting bibliography-output (the contents of the .bbl file)
% must be pasted back into this file before submission.
%
% \bibliographystyle{pss}
% \bibliography{pss-demo}

\begin{thebibliography}{[1]}

\bibitem{Moriya} T. Moriya and T. Takimoto, J. Phys. Soc. Jpn \textbf{64}, 960 (1995). 
\bibitem{Hertz} J.A. Hertz, Phys. Rev. B \textbf{14}, 1165 (1976).
\bibitem{Millis} A.J. Millis, Phys. Rev. B \textbf{48}, 7183 (1993).
\bibitem{Coleman} P. Coleman et al., J. Phys. Condens. Matter \textbf{13}, R723 (2001).
\bibitem{Kambe} S.Kambe et al., J. Phys. Soc. Japan \textbf{10}, 3295 (1996). 
\bibitem{Knafo2} W. Knafo, et al., Phys. Rev. B \textbf{70}, 174401 (2004).
\bibitem{Kadowaki} H. Kadowaki et al., Phys. Rev. Lett.  \textbf{96}, 016401 (2006).
\bibitem{Raymond} S. Raymond et al., J. Low Temp. Phys. \textbf{147}, 215 (2007).
 \bibitem{toto} S. Raymond et al., J. Phys. Condens. Matter \textbf{13}, 8303 (2001).
 \bibitem{Regnault} L.P. Regnault et al., J. Magn. Magn. Mat. \textbf{90-91}, 398 (1990).
 \bibitem{Jacoud} J.L. Jacoud et al., J. Magn. Magn. Mat. \textbf{108}, 131 (1992).
 \bibitem{Mignot}J.-M. Mignot et al., Physica B \textbf{163}, 611 (1990).
 \bibitem{Knafo1}W. Knafo, et al., to be published in Nature Physics 2009.
 \bibitem{Fisher} R.A. Fisher et al., J. Low Temp. Phys. \textbf{84}, 49 (1991).
 \bibitem{Jacoud2} J.L. Jacoud, PhD thesis, Grenoble 1991.
 
\end{thebibliography}
%
% Replace the following example bibliography with your references
% before submission:

\end{document}